\documentclass{ws-procs975x65}

\def\beq{\begin{equation}}
\def\eeq{\end{equation}}

\begin{document}

\title{Pulsars as probes of gravity and fundamental physics}

\author{Michael Kramer$^*$}

\address{Max-Planck-Institut f\"ur Radioastronomie\\
Auf dem H\"ugel 69, 53121 Bonn, Germany\\
$^*$E-mail: mkramer@mpifr-bonn.mpg.de}

\address{Jodrell Bank Centre for Astrophysics\\
University of Manchester, Oxford Road, Manchester M13 9PL, UK}

\begin{abstract}
  Radio-loud neutron stars known as pulsars allow a wide range of
  experimental tests for fundamental physics, ranging from the study
  of super-dense matter to tests of general relativity and its
  alternatives. As a result, pulsars provide strong-field tests of
  gravity, they allow for the direct detection of gravitational waves
  in a 'pulsar timing array', and they promise the future study of
  black hole properties. This contribution  gives an
  overview of the on-going experiments and recent results.
\end{abstract}


\bodymatter


\section{Introduction}

With the direct detection of gravitational waves using Earth-bound
detectors, an enormous milestone in testing gravity has been
achieved.\cite{aaa+16} Even though we had indisputable evidence that
gravitational waves exist from precision pulsar
observations,\cite{ht75a,wnt10,ksm+06} a direct detection on Earth is
important to open up the new field of gravitational wave astronomy.
But even in light of this massive achievement that we can all
celebrate exactly 100 years after Einstein presented his general
theory of relativity (GR), there is still a role of radio astronomy
and pulsar observations in our quest to understand gravity and
fundamental physics. This review will highlight some of the areas
where radio pulsar observations continue to be crucial. It will also
describe how ``Pulsar Timing Array'' (PTA) experiments will join the
Earth-bound experiments to detect gravitational waves directly and how
they complement each other. Finally, after the wonderful evidence for
the existence of stellar-mass black holes presented by the LIGO
collaboration, \cite{aaa+16} we present methods how pulsar
observations can be combined with imaging observations to study the
properties of the supermassive black hole in the centre of the Galaxy
in great detail.

\subsection{Fundamental physics tested using radio astronomy}

Radio photons are the least energetic ones used by astronomers. Yet,
their origin is often associated with highly energetic processes,
coming frequently from areas of extreme conditions involving high
energies, high gravitational or magnetic fields. They are also
relatively easy to detect, to multiply and to interfere with each
other. They can carry polarisation information that can easily be
extracted, allowing us to probe for instance the imprint of
intervening magnetic fields, giving access to plasma physics or
geometric information.  And, even though it may take large
telescope structures to detect the weak radio signals, and even though detectors may
need to be spaced across large distances (and perhaps even into
space), in order to get high spatial resolution images, the radio
window of the electromagnetic spectrum is a doorway to an effective
laboratory for fundamental physics: Testing for a possible a variation
of fundamental constants across cosmic time is possible with the
method of molecular spectroscopy, studying emission that originates
from distant quasars. The Cosmic Microwave Background (CMB) is a
signal from the early Universe that is redshifted so much, that it now
appears in the radio.  Observations of water masers emitting powerful
coherent radio emission allow the establishment of an accurate
distance ladder to measure the local expansion of the Universe. These
and some other examples are summarised in Table
\ref{tab:radiotests}. Despite all these examples, overall there is
arguably no other type of object where astronomy overlaps so widely
with fundamental physics as in the case of radio pulsars.

\begin{table}[h]
\tbl{Selected aspects of fundamental physics studied with
radio astronomical techniques. Note that some
solar system tests have better numerical precision but are derived in
weak gravitational field of the Solar System. In contrast, binary
pulsar limits may sometimes be less constraining in precision, but they
are derived for strongly self-gravitating bodies where deviations are generally
expected to be larger. References are given for more information or
further reading. For a
general review see Will (2014), and for pulsar-related limits see Wex
(2014).}
{\scriptsize
\begin{tabular}{p{2cm}p{3cm}p{4cm}l}\toprule
Tested phenomena & Method & Radio astronomy & Ref.\\
\noalign{\medskip}
\hline
\hline
\noalign{\medskip}
\multicolumn{2}{l}{Variation of fundamental constants:}  & & \\
\noalign{\smallskip}
 Fine structure constant ($e^2/\hbar c$) & Clock comparison, radio
active decays, limit depending on redshift, $< \sim 10^{-16}$ yr$^{-1}$ &
Quasar spectra, $< 10^{-16}$ yr$^{-1}$ & \refcite{wil14} \\
\noalign{\smallskip}
 e-p mass ratio & Clock comparison, $<3.3 \times 10^{-15}$ yr$^{-1}$ &
Quasar spectra, $<3 \times 10^{-15}$ yr$^{-1}$ & \refcite{blc+08},\refcite{ipv+05} \\
\noalign{\smallskip}
 Gravitational constant, $\dot{G}/G$ & Lunar Laser Ranging (LLR),  $(-0.7\pm
3.8) \times 10^{-13}$ yr$^{-1}$ & Binary pulsars, $(-0.6\pm 3.2) \times
10^{-12}$ yr $^{-1}$ & \refcite{hmb10,fwe+12,wex14}  \\
\noalign{\medskip}
\hline
\noalign{\medskip}
 Universality of free fall: & LLR, Nordvedt parm. $|\eta_N| = (4.4 \pm
4.5) \times 10^{-4}$ &
Binary Pulsars, $\Delta < 5.6\times 10^{-3}$  &
\refcite{wil14,sfl+05,wex14} \\
\noalign{\medskip}
\hline
\noalign{\medskip}
\multicolumn{2}{l}{Universal preferred frame for gravity:} & see Table
2 &  \\
\noalign{\medskip}
\hline
\noalign{\medskip}
\multicolumn{2}{l}{PPN parameters and related phenomena:} & see Table
2 & \\
\noalign{\medskip}
\hline
\noalign{\medskip}
\multicolumn{2}{l}{Gravitational wave properties:}  &  Binary pulsars & \refcite{wex14} \\
\noalign{\smallskip}
Verification of GR’s quadrupole formula & & Double Pulsar,
$<\times 10^{-3}$ & \refcite{ks+16} \\
\noalign{\smallskip}
Constraints on dipolar radiation &  & PSR-WD systems,
$(\alpha_A - \alpha_B)^2 < 4 \times 10^{-6}$ & \refcite{fwe+12,afw+13}  \\
\noalign{\medskip}
\hline
\noalign{\medskip}
Geodetic precession & Gravity Probe B, 0.3\% &
PSR B1913+16; Double Pulsar, 13\%; PSR B1534+12, 17\% &
\refcite{edp+11,kra98,bkk+08} \\
\noalign{\medskip}
\hline
\noalign{\medskip}
Equation-of-State & e.g. thermal emission from X-ray binaries & fast
spinning pulsars; massive neutron stars &
\refcite{ls14,hrs+06,dpr+10,afw+13} \\
\noalign{\medskip}
\hline
\noalign{\medskip}
Cosmology & e.g. Supernova distances & Cosmic Microwave Background &
this conference \\
\noalign{\medskip}
\botrule
\end{tabular}}\label{tab:radiotests}
\end{table}

\section{A Simple and Clean Experiment: Pulsars and their Timing}

Pulsars are extremely useful tools for the study of many aspects of
fundamental physics. This is for several reasons: firstly, as neutron
stars they are very compact, strongly-selfgravitating bodies. With
masses of up to $2M_\odot$ (see Section \ref{sec:0348}) concentrated
on a radius of about 10-12 km, or so, they represent the densest
observable matter in the Universe. Understanding the properties of
matter under extreme pressure, i.e. the Equation-of-State, requires
the study of neutron star material. Secondly, pulsars rotate with spin
frequencies up to $\sim700$ Hz, storing an amount of rotational energy, that
makes especially the fast spinning ``millisecond pulsars'' massive stable
flywheels in space. Thirdly, pulsars emit a collimated beam of
coherent radio light along their magnetic axis, making them cosmic
lighthouses, the stable rotation of which can be studied by ``timing''
the arrival times here on Earth. In essence, physicists have been
gifted with a natural precise clock that is attached to a strongly
self-gravitating body. Especially if the pulsar has another compact
binary companion, it can be used to study the effects on the
surrounding spacetime, making it an ideal tool to study gravity under
strong-field conditions.

\subsection{The method}

In pulsar timing, the arrival time of the pulse of a pular is measured
and recorded on Earth. By a comparison of the times measured on Earth,
transferred to the emission time in the pulsar frame, each single
rotation of the pulsar can be tracked and effects influencing the
signal during its propagation through spacetime (at the pulsar and the
solar system) and the interstellar medium can be studied with great
precision. The precision is achieved by describing the pulsar rotation
phase-coherently (i.e.~coherent in rotational phase, so that not a
single rotation is missed). Consequently, all parameters describing
the pulsar (spin-down, astrometry, binary motion, relativistic
effects) included in the ``timing model'' can usually be measured with
a high precision that continues to improve with time. Details of the
procedure can be found, for instance, in Ref.~\refcite{lk05}.

\subsection{The laboratories}

While isolated pulsars also provide insight in a number of aspects of gravity
(e.g.~preferred frame and position effects), the majority of results are obtained
with binary pulsars. The nature of companions represents all possible
outcomes: main-sequence stars, evolved stars, planets, white dwarfs,
other neutron stars and pulsars.\cite{kk16} The only exception is a
current lack of pulsar-black hole systems. These systems should exist,
but they may be rare, and it is still likely that selection effects in
finding compact and fast binary pulsars prevent the successful
discovery so far. Most likely, it is a combination of both, and it
requires a complete census of the Galactic pulsar population (as
planned with the Square Kilometre Array (SKA)\cite{kbc+04,kbk+15}) to
find and calibrate the population synthesis calculations. So far, the
majority of binary systems comprises pulsar-white dwarf systems, but
the number of known so called ``double neutron star'' systems (DNSs)
is increasing steadily (see e.g.~contribution by Freire in this
conference). The results are nevertheless impressive, not only in
variety of measurements, but especially when it comes to precision of
the measurements (Tab.~\ref{tab:precision}).

\begin{table}[hp]
\tbl{Examples of precision measurements using pulsar timing as a
variation  demonstration what is possible today. The digit in bracket indicates
  the uncertainty in the last digit of each value. References are
  cited in the last column.}
{\begin{tabular}{lp{5cm}lp{5cm}}\toprule
Masses: & & \\
Masses of neutron stars: &    $m_1 = 1.4398(2) \,M_\odot$  &  \refcite{wnt10} \\
 &   $m_2 = 1.3886(2) \,M_\odot$ &  \refcite{wnt10} \\
Mass of WD companion:     &          		$0.207(2) \,M_\odot$ & \refcite{hbo06} \\
Mass of millisecond pulsar:	&	$1.67(2) \,M_\odot$ &
\refcite{fbw+11} \\
Main sequence star companion:	&	$1.029(8) \,M_\odot$ & \refcite{fbw+11}\\
Mass of Jupiter and moons:               &
$9.547921(2) \times 10^{-4} \,M_\odot$ &    \refcite{chm+10} \\
\noalign{\medskip}
Spin parameters: & & \\
Period:	&	5.757451924362137(2) ms  & \refcite{vbv+08}  \\
\noalign{\medskip}
Orbital parameters: & & \\
Period:		&		$0.102251562479(8)$ day &   \refcite{ks+16} \\
Eccentricity:	&		$3.5 (1.1) \times 10^{-7}$ &
\refcite{fwe+12} \\
\noalign{\medskip}
Astrometry:& & \\
Distance:      &         	 		$157(1)$ pc	& \refcite{vbv+08}\\
Proper motion: &  		    	$140.915(1)$ mas yr$^{-1}$ & 	\refcite{vbv+08}\\
\noalign{\medskip}
Tests of general relativity: & & \\
Periastron advance:	&		$4.226598(5)$ deg yr$^{-1}$ & 	\refcite{wnt10}\\
Shrinkage due to GW emission: 	&                      $7.152(8)$
mm/day & 	\refcite{ks+16} \\
GR validity (obs/exp):	&	$1.0000(5)$ & \refcite{ks+16} \\
Constancy of grav. Constant, $\dot{G}/G$: &           $-0.6(1.6) \times
10^{-12}$ yr$^{-1}$ &             \refcite{fwe+12}\\
\botrule
\end{tabular}}
\label{tab:precision}
\end{table}

\section{Pulsars as Gravitational Wave Detectors}

The idea of using pulsars not only as sources of GWs
in binary systems, but to also use them as GW detectors is 40 years
old. The idea is simple, but as for LIGO, the realisation is a
challenge - but also one that should be mastered eventually. If a
low-frequency GW distorts the local space-time near Earth, it should
be visible  as an red noise signal in the pulsar timing
residuals. In order to recognise it as an astrophysical signal, we
need to distinguish it from other sources of noise. This is possible
since distortions of the local space-time affect all timed pulsars,
not in the same way, but as described by the polarisation
characteristics of the GW. For an (expected) quadurpolar polarisation,
pulsars in the same or opposite direction on the sky should show a
positive correlation in their arrival time (i.e.~the pulses arrive
simultaneously too early or too late), while pulsars at right angles
on the sky, should be anti-correlated. The exact relationship for a
stochastic background of GWs is expressed in the
``Helling-and-Downs'' curve (for quadrupolar polarisation, for other
polarisation modes see below) as first computed in
Ref.~\refcite{hd83}.

In other words, it requires an ensemble of pulsars to make a reliable
detection. The frequency range that this ``Pulsar Timing Array'' (PTA) is
sensitive to depends both the cadence of the observations and the
length of the data set. With a typical cadence of days to weeks, and a
time-baseline of nearly 20 years or more (see e.g.~the recent
results of the European Pulsar Timing Array, EPTA, with data sets
spanning up to 18 years,\cite{dcl+16}) this corresponds to a frequency range
from (18 years)$^{-1} =  1.8$nHz  to (1 week)$^{-1}=  1600$ nHz. (NB: The
``high'' frequency end is less probed currently, as high cadence
observations were started only relatively recently for most sources).
The resulting shape of the sensitivity curve plotted into a diagram of
characteristic strain vs. GW frequency is wedge-like, with the sharp
low-frequency end given by the data set length. At the frequency of
1 year$^{-1}$, the PTA are insensitive, as any GW signal would be
absorbed for a fit to the astrometric parameters, measurable due to
the Earth movement about the Sun.

\subsection{Status of the PTA efforts}

There are three major PTA experiments in the world at the
moment. There is the European Pulsar Timing Array (EPTA),\cite{kc13}
the North American Gravitational Wave Observatory (NANOGrav)\cite{mcl13}
and the Parkes Pulsar Timing Array (PPTA).\cite{hob13} All of these
experiments conduct regular multi-frequency timing observations of 20
to 40 millisecond pulsars. The aim of these efforts is to detect GWs
in the nHz-regime. The strongest signal that can be expected is the
superposition of signals emitted by super-massive black holes binaries
(SMBHBs). In the simplest form, one would expect a power-law signal
with a spectral index of $f^{-2/3}$ and a characteristic strain
amplitude at a frequency of (1 year)$^{-1}$ of about $10^{-15}$. In
reality the shape of the spectrum depends on astrophysical processes
during galaxy mergers (for a review see, for instance,
Ref.~\refcite{ses16}), demonstrating that an eventual detection of the
signal is not only important for GW research, but also for studies of
the physics of galaxy formation and mergers. So far, only upper limits
on the stochastic signal could be obtained.  With similar and
complementary capabilities the results of all PTA are comparable,
producing limits that are all comparable within factors 2 to 3 (see
Refs.~\refcite{srl+15,ltm+15,abb+16}). In order to increase the sensitivity and
frequency coverage of the PTAs the three experiments also pool their
data within the International Pulsar Timing Array (IPTA). It can be
expected that the current results\cite{vlh+16} will improve
considerably in the future. In particular the addition of the most
recent and hence most accurate data will go a long way towards a
detection that would complement the Earth-bound (and later
space-bound) window with very low nHz-frequencies. In particular the
addition of results from the Large European Array for Pulsars (LEAP)
\cite{kc13,bjk+16} promise to boost the overall EPTA and IPTA
capabilities. See also the contribution by Perrodin and others for
more details.

\subsection{PTA science beyond detection}

Like in the case of LIGO, a PTA detection of GWs would only be the
first step towards physics and astrophysics enabled by GW astronomy.
Understanding the astrophysics of SMBHB mergers is only one obvious
topic to be better understood by PTA signals. A high-precision
measurement of the Hellings-Downs curve enabled by the SKA, also
allows to measure or constrain the properties of gravitational
waves. It is easy to see that the shape of the curve is affected by
deviations from a quadrupolar nature of the GWs, or a deviation of the
graviton spin from 2. Indeed, in alternative theories, up to six
polarisation modes are possible, that can be explored and constrained by
very sensitive PTA observations. Similarly, but perhaps less obvious,
is the fact that a non-zero mass of the graviton would introduce a
cut-off frequency beyond which GW propagation in the Universe is not
possible. This also modifies the Helling-Downs curve and should be
detectable in future SKA observations.
\cite{ljp08,ljp+10,kbc+04,jhm+15}

\section{Constraining PPN Parameters}

Tests in the weak-field limit, as for instance conducted in the solar
system, can be performed within the framework of the Parameterised
Post-Newtonian (PPN) formalism.\cite{wn72} Here, a particular effect
can be associated with a particular PPN parameter, which assumes a
certain value in GR, but may have different values in alternative
theories of gravity. Table \ref{tab:PPNpar}, adapted from a similar
table in Ref.~\refcite{wil06}, summarises the PPN parameters and their
values in which a general theory of gravity can differ from GR at 1PN level.
While they are used to describe theories in the limit in
which the gravitational field is weak and generated by objects moving
slowly compared to the speed of light, it is possible to write down
``strong-field equivalents'', which can then be tested with
pulsars. Again, Table \ref{tab:PPNpar} summarised the current best
limits, where indeed the majority is now constrained by pulsars (see
Ref.~\refcite{wex14}) for details).

\begin{table}[h]
\tbl{Best limits for the parameters in the PPN formalism.
Note that 6 of the 9 independent PPN parameters are best
constrained by radio astronomical techniques. Especially, five
of them are derived from pulsar observations.
Adapted from Will (2014) but see also Wex (2014) for details.}
{\scriptsize
\begin{tabular}{lp{3cm}lrp{2.5cm}}\toprule
Par. & Meaning & Method  & Limit  & Remark/Ref.\\
\noalign{\medskip}
\hline
\hline
\noalign{\medskip}
$\gamma-1$ &  How much space-curvature produced by unit rest mass? & time delay &  $2.3 \times 10^{-5}$ & Cassini tracking/\refcite{wil14} \\
  & & light deflection &  $2\times 10^{-4}$ & VLBI/\refcite{wil14} \\
\noalign{\medskip}
\hline
\noalign{\medskip}
$\beta -1$ & How much ``non-linearity'' in the superposition law for
gravity? & perihelion shift & $8 \times 10^{-5}$ &  using
$J_{2\odot}=(2.2\pm0.1)\times 10^{-7}$/\refcite{wil14} \\
 &  & Nordtvedt effect &  $2.3\times 10^{-4}$ &  $\eta_{N}  =4\beta-\gamma-3$ assumed \\
\noalign{\medskip}
\hline
\noalign{\medskip}
$ \xi$ & Preferred-location effects? & spin precession & $4\times 10^{-9}$ & Isolated MSPs/\refcite{sw13} \\
\noalign{\medskip}
\hline
\noalign{\medskip}
$\alpha_1$ & Preferred-frame effects? & orbital polarisation  & $4\times 10^{-5}$ & PSR-WD, PSR J1738+0333/\refcite{sw12} \\
\noalign{\smallskip}
$\alpha_2$ &  & spin precession & $2 \times 10^{-9}$ & Using isolated MSPs/\refcite{sck+13} \\
\noalign{\smallskip}
$\alpha_3$ & & orbital polarisation  & $5.5\times 10^{-20}$ & Using ensemble
of MSPs/\refcite{gsf+11} \\
\noalign{\medskip}
\hline
\noalign{\medskip}
$\zeta_1$ & Violation of conversation of total momentum? & Combining PPN bounds &
$2\times 10^{-2}$ & \refcite{wil14} \\
\noalign{\smallskip}
$\zeta_2$ & & binary acceleration & $4\times 10^{-5}$ & Using $\ddot{P}$ for
PSR B1913+16/\refcite{wil14} \\
\noalign{\smallskip}
$\zeta_3$ & & Newton's 3rd law & $10^{-8}$ & lunar acceleration/\refcite{wil14} \\
\noalign{\smallskip}
$\zeta_4$ &  & not independent  &  &  $6\zeta_4 = 3\alpha_3 + 2\zeta_1 − 3\zeta_3$  \\
\noalign{\medskip}
\botrule
\end{tabular}}\label{tab:PPNpar}
\end{table}

\section{Binary Pulsars}

Binary pulsars in compact, ideally eccentric, orbits, may show a number
of relativistic effects that influence the pulse arrival times in a
variety of ways. They manifest themselves as deviations from the
classical Keplerian motion of these ``relativistic binaries'', and they
can be described by theory-independent corrections to the binary
motion, so called ``Post-Keplerian (PK) Parameters'' as introduced
by Refs.~\refcite{dam88,dt92}. The PK parameters depend on the
well-measured Keplerian parameters and the {\em a priori} unknown
masses of the binary components, whereas the functional dependence on
them is determined by a given theory of gravity. With the measurement of two PK
parameters, the two masses can be determined, assuming the
particular theory. The theory can be tested for consistency if one or more
additional PK parameters can be measured. The measured value can be
compared with the one determined using the functional dependence and
the calculated masses. If the values disagree, the chosen theory is
falsified.

\subsection{The Hulse-Taylor pulsar}

The first system, where such a described test was possible, was the
first binary pulsar discovered by Hulse and Taylor, PSR B1913+16
(Ref.~\refcite{ht75a}). The system allowed the measurement of three PK
parameters.

The easiest to be measured for binary pulsars (unless the orbit is
circular) is the advance of periastron, which in GR is given by
\begin{equation}
\dot{\omega} = 3 T_\odot^{2/3} \; \left( \frac{P_{\rm b}}{2\pi} \right)^{-5/3} \;
               \frac{1}{1-e^2} \; (m_p +
               m_c)^{2/3}. \label{omegadot}
\end{equation}
Here, $T_\odot=GM_\odot/c^3=4.925490947 \mu$s is a
constant, $P_{\rm b}$ the orbital period, $e$ the
eccentricity, and $m_p$ and $m_c$ the masses of the pulsar and its
companion. See Ref.~\refcite{lk05} for further details.

The second parameter describes the effects of time dilation as the
pulsar moves in its elliptical orbit at varying distances from the
companion and with varying speeds. In GR, the observed amplitude of
the integrated effect is related to the Keplerian parameters and the
masses as
\begin{equation}
\gamma_E  = T_\odot^{2/3}  \; \left( \frac{P_{\rm b}}{2\pi} \right)^{1/3} \;
              e\frac{m_c(m_p+2m_c)}{(m_p+m_c)^{4/3}}.
\end{equation}
By measuring $\dot{\omega}$ and  $\gamma_E$, one can determine the
masses assuming GR,  $m_p = 1.4398 \pm 0.0002 \,M_\odot$ and $m_c = 1.3886 \pm 0.0002
\,M_\odot$ (Ref.~\refcite{wnt10}).

With time, it was possible to measure the shrinkage of the orbit due
to the emission of gravitational waves, which manifests itself in a
shift of periastron time to earlier times. This allows to determine
the change in orbital period, which in GR depends on
\begin{equation}
\dot{P}_{\rm b} = -\frac{192\pi}{5} T_\odot^{5/3} \; \left(
  \frac{P_{\rm b}}{2\pi} \right)^{-5/3} \;
               \frac{\left(1 +\frac{73}{24}e^2 + \frac{37}{96}e^4 \right)}{(1-e^2)^{7/2}} \;
               \frac{m_p m_c}{(m_p + m_c)^{1/3}}.
\end{equation}
The predicted value using this relationship agrees with the observed
value, however, only if a correction for a relative acceleration
between the pulsar and the solar system barycentre is taken into
account. As the pulsar is located about 7-8 kpc away from Earth, it
experiences a different acceleration in the Galactic gravitational
potential than the solar system (see e.g. ~Ref.~\refcite{lk05}). The
precision of our knowledge to correct for this effect eventually
limits our ability to compare the GR prediction to the observed
value. Nevertheless, the agreement of observations and prediction,
today within a 0.2\% (systematic) uncertainty (Ref.~\refcite{wnt10}),
represented the first evidence for the existence of gravitational
waves.

\subsection{The Double Pulsar}

Today we know many more binary pulsars where we can detect
gravitational wave emission. In the particular case of the Double
Pulsar, the measurement uncertainties are not only more precise, but
also the systematic uncertainties are much smaller, as the system is
much more nearby. However, what makes the system particularly unique
is the fact that the system consists of two active radio
pulsars.\footnote{As we describe later, one of them is temporarily not
  visible due to the effects of relativistic spin precession.}

One pulsar is mildly recycled with a period of 23 ms (named ``A''), while the other
pulsar is young with a period of 2.8 s (named ``B''). Both orbit the common centre
of mass in only 147-min with orbital velocities of 1 Million km per
hour. Being also mildly eccentric ($e=0.09$), the system is an ideal
laboratory to study gravitational physics and fundamental physics in
general. A detailed account of the exploitation for gravitational
physics has been given, for instance, by Refs.~\refcite{ksm+06,ks08,kw09}.
An update on those results is in preparation (Ref.~\refcite{ks+16}),
 with the largest improvement undoubtedly given by a
large increase in precision when measuring the orbital decay. Not
even ten years after the discovery of the system, the Double Pulsar
provides the best test for the accuracy of the gravitational
quadrupole emission prediction by GR far below the 0.1\% level.

The fact that the size of both pulsar orbits, $x_A$ and $x_B$ could be
measured, it is possible to determine the ratio of the masses pulsars
A and B, $R=m_A/m_B = x_B/x_A$ in a theory-independent fashion (at
least to the 1PN level, Ref.~\refcite{dd86}). In addition to the
periastron advance, $\dot{\omega}$, and the gravitational redshift
parameter, $\gamma$, also a Shapiro delay could be measured, which
provides two PK parameters, i.e. the ``shape'' $s$ and ``range'' $r$,
given in GR by
\begin{equation}
s = T_\odot^{-1/3} \; \left( \frac{P_{\rm b}}{2\pi} \right)^{-2/3} \; x \;
              \frac{(m_A + m_B)^{2/3}}{m_B} \label{eqn:s},
\end{equation}
and
\begin{equation}
r = T_\odot m_B, \label{eqn:r}\\
\end{equation}
The shape parameter relates to the orbital inclination angle $i$, via
$s=\sin(i)$, and is measured to be very close to  an
almost perfect edge-on configuration,\cite{ks+16} which is
also reflected in the observation of 30-s long eclipse of
pulsar A due to the blocking magnetosphere of B. This eclipse is not
perfect, but leads to the modulation of the registered emission of A,
as the torus (unlike a disk) does not block the light
continuously. This  leaves an imprint in A's observed flux density
with a characteristic pattern, showing the emission of A every
half-period of B first and later only every full-period of B, before
the eclipse is over. This pattern depends on the relative orientation
of B's spin axis and therefore changes slowly due to geodetic
precession. A global fit to these eclipse pattern allowed us to
determine the precession rate of pulsar B,\cite{bkk+08} which in GR
is given by
\begin{equation}
\label{eqn:om}
\Omega_{\rm B}  =  T_\odot^{2/3} \left( \frac{2\pi}{P_{\rm b}}\right)^{5/3}
 \frac{m_{\rm A}(4m_{\rm B}+3m_{\rm A})}{2(m_{\rm B}+m_{\rm A})^{4/3}}
 \frac{1}{1-e^2}
\end{equation}
The observed value agrees with GR's prediction within the uncertainty
of 13\%. The precision of this value will improve with time, even
though pulsar B is not visible at the moment. Indeed, the same precession that
modulates the eclipse pattern has moved B's beam out of our
line-of-sight, so that it momentarily misses Earth.\cite{pmk+10}

In principle, one can write down Eqn.~\ref{eqn:om} also for pulsar
A. However, geodetic precession is not observed for A. It would
manifest itself in a change of the pulse profile (as it had done for
the Hulse-Taylor pulsar\cite{wrt89,kra98} and pulsar B,\cite{bpm+05})
but the profile is observed to be extremely stable.\cite{fsk+13} This
leads to the conclusion that the spin axis of A is nearly perfectly
aligned (within less than 3 deg) with the orbital
angular momentum vector. This in turn is evidence for a formation of B
in a low-kick supernova  (Ref.~\refcite{fsk+13}).

\section{Constraining Alternative Theories}

Even though GR has been very successful in describing all experiments
in the solar system, binary pulsar tests, and now also the observed
LIGO detection,\cite{aaa+16} phenomena like ``dark matter'' or``dark energy'' may
suggest deviations from GR under certain conditions or on certain
scales. Alternatives to GR are therefore discussed and need to be
confronted with experimental data. A particular sensitive test
criterion is if the theory is able to make a statement
(i.e.~prediction!) about the
existence and type of gravitational waves emitted by binary pulsars.

A class of alternative theories where intensive work has made this possible, are
scalar-tensor theories as discussed and demonstrated
in a series of works (e.g.~Ref.~\refcite{ear75,wg89,de96}).
For corresponding tests, the choice of a double neutron star system is
not necessarily ideal, as the difference in scalar coupling, (that would be
relevant, for instance, for the emission of gravitational {\em dipole}
radiation) is small. The ideal laboratory would be a pulsar orbiting a
black hole, as the black hole would have zero scalar charge.
But also compact pulsar-white dwarf systems can be useful in
providing constraints on alternative theories of gravity that are
equally good or even better than solar system limits
(Ref.~\refcite{fwe+12}).

The currently best example for such a system
PSR J1738+0333, a 5.85-ms pulsar in a practically circular
8.5-h orbit with a low-mass white dwarf companion.\cite{fwe+12}
The determination of the intrinsic orbital decay due to gravitational
wave emission shows an agreement with the prediction of GR, hence
introducing a tight upper limit on dipolar gravitational wave
emission. This can be translated into stringent constraint on
general scalar-tensor theories of gravity. The new bounds are more
stringent than the best current Solar system limits over most of the
parameter space, and constrain the matter-scalar coupling constant
$\alpha_0^2$ to be below the $10^{-5}$ level. For the special case of
the Jordan-Fierz-Brans-Dicke theory, the authors obtain a one-sigma
bound of $\alpha_0^2< 2 \times 10^{-5}$, which is within a factor of
two of the Solar-System Cassini limit.

\label{sec:0348}

The relativistic pulsar-white dwarf system PSR J0348+0432 is on course
to provide better limits in the near future. The system can be studied
both in the optical (via the white dwarf emission) as well as the
radio (via the pulsar).\cite{afw+13}  On one hand, this allows to
measure the mass of the neutron star, showing that it has a
record-braking value of $2.01\pm0.04M_\odot$! This is not only the
most massive neutron star known (at least with reliable precision),
but with the also clearly measured gravitational wave damping of the
orbit, the system is a sensitive laboratory of a previously untested
part of the strong-field gravity regime. Thus far, the observed orbital decay
agrees with GR, supporting its validity even for the extreme
conditions present in the system.\cite{afw+13}

The recent discovery of PSR J0337+1715, a millisecond pulsar in a
hierarchical triple system with two other white dwarfs offers another
exciting laboratory to study in particular possible violations of the
Strong Equivalence Principle.\cite{rsa+14} As the inner pair of pulsar
and white dwarf orbit every 1.63 days in the gravitational field of an
outer white dwarf (with an orbital period of 327 days), the impact of
this external gravitational field on the inner orbital motion can be
tested, as described previously in Ref.~\refcite{fkw12}.  In the near
future we expect excellent limits on scalar-tensor-theories using this
system (Ref.~\refcite{bbc+15}).  See also the contribution by
Archibald in this conference for more details.

\begin{figure}[htb!]
\begin{center}
\includegraphics[width=12cm]{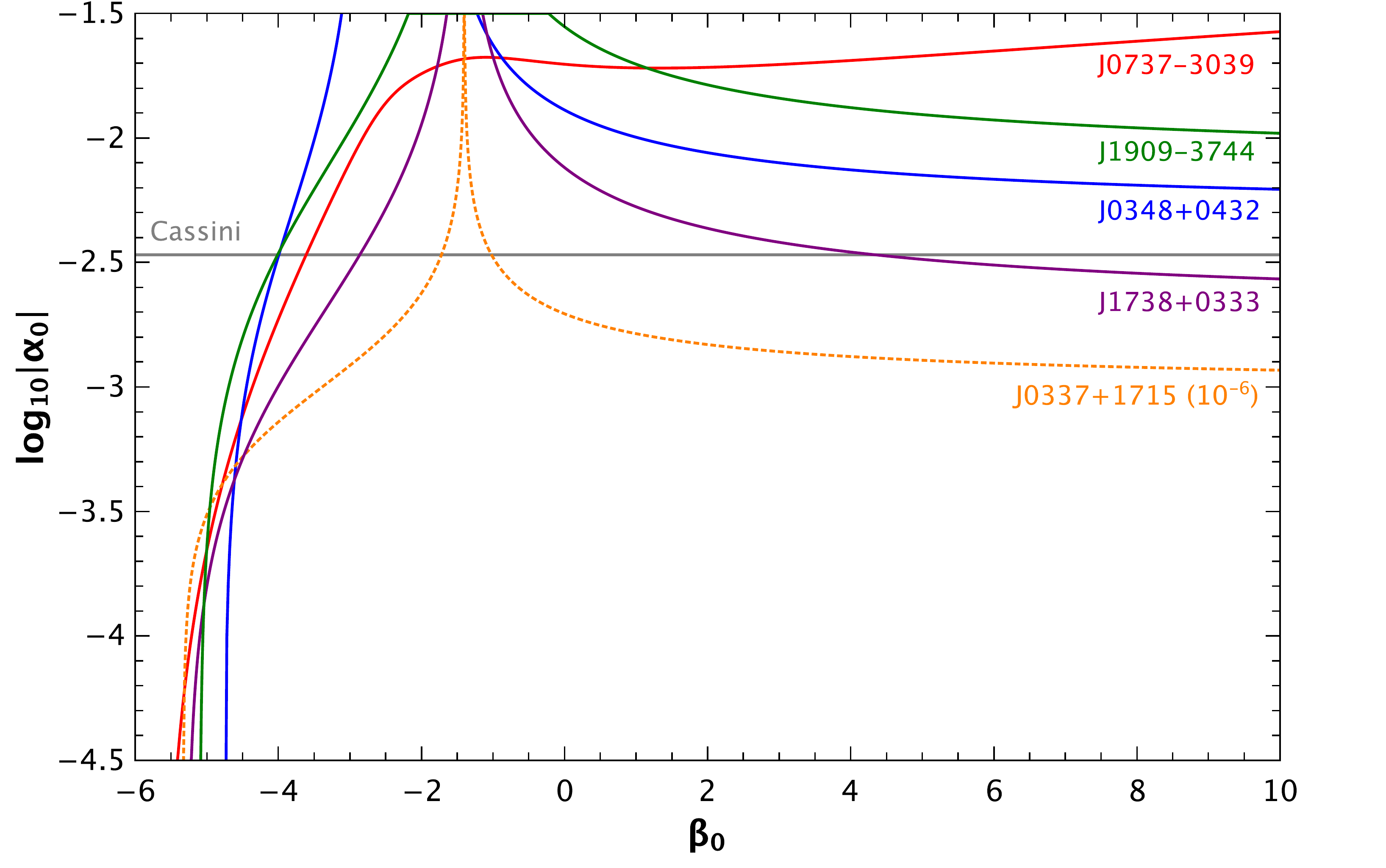}
\end{center}
\caption{Conservative limits on quadratic mono-scalar-tensor gravity (cf.\
Ref.~\protect\refcite{de96}), from the Cassini spacecraft 
(Ref.~\protect\refcite{bit03}) and different radio pulsars. The area to the left 
and the top of a curve is excluded by the corresponding test. The curve for the 
Double Pulsar (J0737$-$3039) corresponds to a 0.1\% gravitational wave test. As 
indicated in the text, by now the Double Pulsar has already well surpassed 
this limit. The dotted line shows a $10^{-6}$ test of 
the universality of free fall with the pulsar in the triple system (comparable 
to the preliminary limits presented by Anne Archibald at MG14). The 
limits for pulsars account for the uncertainty in the neutron-star equation of 
state (EoS), by excluding only those parts in the $\alpha_0$--$\beta_0$ theory 
plane which fail the gravity test with the corresponding pulsar for all EoSs, 
chosen from a large set of EoSs ranging from very soft to very stiff and 
being compatible with the mass of PSR J0348+0342. Jordan-Fierz-Brans-Dicke 
gravity corresponds to the line $\beta_0 = 0$. Figure provided by N.~Wex.}
\label{fig:stg}
\end{figure}

Figure~\ref{fig:stg} summarizes our present constraints on a specific class of 
scalar-tensor gravity, which also accounts for our current imperfect knowledge 
of the equation of state for neutron-star matter. A summary of selected further 
tests of alternative theories of gravity using pulsars is given in 
Table~\ref{tab:theories}.


\begin{table}[h]
\tbl{Constraining specific (classes of) gravity theories using radio pulsars.
See text and also Wex (2014) for more details.}
{\footnotesize
\begin{tabular}{@{}lp{7cm}p{1cm}@{}}\toprule
Theory (class)  & Method & Ref. \\
\noalign{\medskip}
\hline
\noalign{\medskip}
\underline{Scalar-tensor gravity:} & & \\
\noalign{\smallskip}
Jordan-Fierz-Brans-Dicke &  limits by PSR J1738+0333 and PSR
J0348+0432, comparable to best Solar system test (Cassini)  & \refcite{fwe+12,bbc+15}  \\
\noalign{\smallskip}
Quadratic scalar-tensor gravity &  for $\beta_0 < -3$ and $\beta_0 >
+5$ best limits from PSR-WD systems, in particular PSR J1738+0333 and
PSR J0348+0432 &  \refcite{fwe+12,bbc+15}   \\
\noalign{\smallskip}
Massive Brans-Dicke & for $m_\varphi \sim 10^{-16}$ eV: PSR J1141$-$6545 &  \refcite{abwz12}\\
\noalign{\medskip}
\hline
\noalign{\medskip}
\underline{Vector-tensor gravity:} & & \\
\noalign{\smallskip}
Einstein-\AE{}ther & combination of pulsars (PSR J1141$-$6545,
PSR J0348+0432, PSR J0737$-$3039, PSR J1738+0333) &  \refcite{ybby14}\\
\noalign{\smallskip}
Ho{\v r}ava gravity & combination of pulsars (see above) & \refcite{ybby14}   \\
\noalign{\medskip}
\hline
\noalign{\medskip}
\underline{TeVeS and TeVeS-like theories:} & & \\
\noalign{\smallskip}
Bekenstein’s TeVeS & excluded using Double Pulsar  &  \refcite{ks+16} \\
\noalign{\smallskip}
TeVeS-like theories & excluded using PSR 1738+0333 & \refcite{fwe+12} \\
\noalign{\medskip}
\botrule
\end{tabular}}\label{tab:theories}
\end{table}

\section{Pulsar-Black Hole Systems}

As described above, in many respects a pulsar-black hole system would
be the ideal system to test theories of gravity. The pulsar could be
as probe that can be used to trace the motion in the spacetime of the
black hole. The first basic recipe of how to extract the black hole
spin and quadrupole moment from such a system was presented by
Ref.~\refcite{wk99}. It was shown that by following the orbital motion
and by measuring the higher order time derivatives of the the
projected semi-major axis, $x$, and the periastron angle, $\omega$,
the ``Cosmic Censorship Conjecture'' and the ``No-hair theorem'' could
be tested. This formed the basis of the science case for the
corresponding SKA Key Science case (Ref.~\refcite{kbc+04}). It was
later shown that measuring the spin for stellar-mass black holes was
possible, but that extracting the quadrupole moment was difficult even
with the SKA.\cite{lewk14} Nevertheless, a stellar-mass black-hole -
pulsar system would still provide a superb probe for theories of
gravity (see e.g.~Refs.~\refcite{wex14,ysy16}) and so the hunt to find such a
system continues. One place where we expect pulsars to orbit a black
hole is the centre of our Galaxy. The large mass of Sgr A* would also
make the measurement of the black hole parameters, including the
quadrupole moment, possible, as shown by Ref.~\refcite{lwk+12}. As
shown recently by Ref.~\refcite{pwk16}, it is not even necessary to
measure the complete full orbit to determine the BH parameters, but
measuring the pulsar around periapsis is sufficient. This implies that
certain perturbations of the orbit away of periapsis will not prevent
the successful outcome of such an experiment.

\subsection{Studying the super-massive black hole in the Galactic Centre}

As shown in Ref~\refcite{lwk+12}, a slow, normal pulsar in an appropriate orbit
would be sufficient to in principle measure the mass of SGR A* with a precision 
of a few $M_\odot$, to test the cosmic censorship conjecture to a
precision of about 0.1\% and to test the no-hair theorem to a
precision of 1\%. This is possible even with a rather modest timing
precision of $100\mu$s due to the large mass of SGR A* and the
measurement of relativistic and classical spin-orbit coupling,
including the detection of frame-dragging. Unlike other methods, Liu
et al.~also developed a method that allows us to test for a possible
contamination of the orbital measurements by nearby stars. Given the
huge rewards for finding and timing pulsars in the Galactic centre,
various efforts have been conducted in the past to survey the inner
Galaxy and the Galactic centre in particular
(e.g., Refs.~\refcite{kkl+00,jkl+06,dcl09,mkfr10}). None of these
efforts has been successful, despite the expectation to find more than
1000 pulsars, including millisecond pulsars (e.g.~Ref.~\refcite{wcc+12})
or even highly eccentric stellar BH-millisecond pulsar systems
\cite{fl11}. In reality, external perturbations of the orbit (e.g.~
due to the presence of other stars) may limit the precision of the
possible measurements. But as shown recently by Ref.~\refcite{pwk16},
the repeated observations of the periapsis is sufficient to extract the
information even in such cases, albeit with reduced precision.

The lack of detection was thought to be understood in terms of severely increased
interstellar scattering due to the highly turbulent medium. Scattering
leads to pulse broadening that cannot be removed by instrumental means
and that renders the source undetectable as a pulsar, in particular if
the scattering time exceeds a pulse period. The scattering time,
however, decreases as a strong function of frequency ($\propto
\nu^{-4}$, see e.g. Ref.~\refcite{lk05}), so that the aforementioned
pulsar searches have been conducted at ever increasing frequencies --
the latest being conducted at around 20 GHz. The difficulty
in finding pulsars at these frequencies is two-fold. On one hand, the
flux density is significantly reduced due to the steep spectra of
pulsars. On the other hand, the reduced dispersion delay, which
usually needs to be removed but also acts as a natural discriminator
between real pulsar signals and man-made radio interference, is making
the verification of a real signal difficult.

The situation changed in 2013, when triggered by a detection of a
periodic X-ray source by SWIFT and NuStar (\refcite{kbk+13,mgz+13}),
Ref.~\refcite{efk+13} were the first to discover a 3.8-s magnetar
using the 100-m Effelsberg radio telescope only 2'' away from Sgr
A*. The magnetar, PSR J1745$-$2900, has the hightest dispersion
measure of any pulsar, is highly polarised and has a rotation measure
that is larger than that of any other source in the Galaxy, apart from
Sgr A*.  The fact that a rare object like a radio-emitting magnetar
may be found in such proximity to Sgr A* suggests that many more
ordinary pulsar may exist.  However, the magnetar also showed a
scattering of emission that was far less than expected for the
Galactic Centre, begging the questions, why many pulsars, should they
exist, were not yet detected after all.  Solving this question is at
the focus of on-going research, and by the time of the next
Marcel-Grossmann meeting, new interesting results may be available.

\section{Pulsars and an Image of Sgr A*}

An international effort is underway to perform a mm-VLBI experiment
known as the ``Event Horizon Telescope'' (EHT) with the aim to image
the shadow of Sgr A* against the background radiation from a hot
accretion disk (e.g.~Ref.~\refcite{fm13}).  With a mass of about
$4.3\times 10^6 M_\odot$ (e.g.~Ref~\refcite{gef+09}), the central BH is
not very large in size compared to those in the centre of other
galaxies, but it is the closest.  The idea is that the precise
measurement of the shadow also leads to an extraction of the black
hole parameters, assuming that the shadow is determined purely by
gravitational effects and that modeling does not depend on
understanding of the accretion flow properties (see
e.g. Ref.~\refcite{fm13} for a recent review and contributions to this
conference).

As shown recently by Ref.~\refcite{pwk16}, the correlated
uncertainties in the measurements of the black hole spin and
quadrupole moment using pulsars are nearly orthogonal to those
obtained from measuring the shape and size of the black hole
shadow. Combining the different types of observations allows one to
assess and quantify systematic biases and uncertainties in each
measurement, which will finally lead to a highly accurate,
quantitative test of the no-hair theorem. This is possible since the
image and the pulsar measurements probe simultaneously the near- and
far-field of SGR A*, promising a unique probe of gravity.

\section{Summary}

We are in a golden age of testing theories of gravity. Direct
observations of GW allow to probe the dynamic strong-field regime, and
allow us to probe the properties of black holes. Such tests probe a
different regime that those of binary pulsars, which complement the
efforts by testing self-field effects in strongly self-gravitating
bodies like neutron stars. Together with direct observations of the
event horizon of black holes, all  experiments are therefore complementary
and together probe a theory-space that will eventually inform us about
the underlying theory of gravity.



\section*{Acknowledgments}

BlackHoleCam is supported by the ERC Synergy Grant BlackHoleCam under grant agreement No.~610058.


\end{document}